\documentclass[useAMS,usenatbib,usegraphicx]{mn2e}
\usepackage{multirow}
\usepackage{url}

\title[The \emph{ASCA}/\emph{INTEGRAL} source AX~J183039--1002]{The nature of the \emph{ASCA}/\emph{INTEGRAL} source AX~J183039--1002: a new Compton-thick AGN?}

\author[L. Bassani et al.]
{L.~Bassani,$^1$\thanks{E-mail address: \texttt{bassani@iasfbo.inaf.it}} 
R. Landi,$^{1}$
R. Campana,$^{2}$
V. A. McBride,$^{3}$
A. J. Dean,$^{3}$
A. J. Bird,$^{3}$\newauthor
D. A. Green,$^{4}$
P. Ubertini,$^{5}$
A. De Rosa$^{5}$ \\
$^1$ INAF/IASF-Bologna, Via P. Gobetti 101, I-40129 Bologna, Italy \\
$^2$ Department of Physics, University of Rome ``La Sapienza'', Piazzale A. Moro 2, I-00185, Rome, Italy \\
$^3$  School of Physics and Astronomy, University of Southampton, Highfield, SO17 1BJ, United Kingdom \\
$^4$ Astrophysics Group, Cavendish Laboratory, 19 J. J. Thomson Ave., Cambridge CB3 0HE, United Kingdom \\
$^5$ INAF/IASF-Roma, Via Fosso del Cavaliere 100, I-00133, Roma, Italy \\
}

\date{Accepted 2009 January 16.  Received 2009 January 15; in original form 2008 October 10.}
\pagerange{\pageref{firstpage}--\pageref{lastpage}} \pubyear{2008}

\begin{document}
\label{firstpage}
\maketitle 

\begin{abstract}
We report on the identification of the X/soft $\gamma$-ray source AX~J183039--1002 detected with \emph{ASCA} and \emph{INTEGRAL}/IBIS. The source, which has an observed 20--100 keV flux of $\sim8.6\times 10^{-11}$ erg cm$^{-2}$ s$^{-1}$, is inside a diffuse radio supernova remnant (SNR) and is spatially coincident with a compact radio source.
We analyzed archival  \emph{Chandra} and \emph{XMM-Newton} observations in order to identify the \emph{ASCA/INTEGRAL} source. 
A point-like \emph{Chandra} X-ray object was found to be positionally coincident with the compact radio source and within the error circle of the \emph{ASCA} and \emph{INTEGRAL} sources. 
Although the association of a compact radio/X-ray source with a radio supernova remnant could be indicative of a pulsar wind nebula (PWN), the \emph{XMM-Newton} X-ray spectrum is compatible with an absorbed, Seyfert-2 like AGN, since it provides evidence for an iron emission line of $\sim$ 1 keV equivalent width; furthermore the X-ray source spectrum is similar to that of other Compton thick AGN where the 
$\la$2 keV data are associated to a warm reflector and the $\ga$10 keV one to a cold reflector.
\end{abstract}


\section{Introduction}

A key strategic objective of the \emph{INTEGRAL} (\emph{International Gamma Ray Astrophysics Laboratory}) mission is a high energy survey of the Galactic plane and 
Centre (Winkler et al. 2003) \emph{in primis} and of all the sky as a by-product of pointing observations. 
This makes use of the unique imaging capability of IBIS (\emph{Imager on Board Integral Satellite}; Ubertini et al. 2003), which allows the detection of 
sources at the mCrab level, with an angular resolution of 12\arcmin\ and a point source location 
accuracy (PSLA) of typically 1\arcmin--3\arcmin\ within a large ($29\degr\times29\degr$) field of view. So far, 
several surveys produced from the IBIS/ISGRI data have been reported, the most complete being those of Bird et al. (2007) and Krivonos et al. (2007); 
both reported around 400 sources down to a flux level of a few mCrab above $\sim$20 keV. 
A significant fraction (25--30$\%$) of the objects in these surveys have no obvious counterpart at other wavelengths and therefore cannot yet be associated with any known class of high energy emitting objects. 
Searching for counterparts of these new sources is of course a primary objective of the survey work but it is made very difficult by the relatively large \emph{INTEGRAL} error boxes. 
Cross correlations with catalogues in other wavebands can be used as a useful tool with which to restrict the positional uncertainty of the objects detected by IBIS and therefore to facilitate the identification process. 
Observations at softer X-ray energies, where the positional accuracy is much better, has already proved an invaluable aid in the identification and classification process (Stephen et al.  2006; Masetti et al. 2008 and references therein), but the use of data at other wavebands, for example the radio, is also useful to help identify peculiar and interesting objects.

In this Letter we report on the unusual nature of the \emph{ASCA}/\emph{INTEGRAL} source AX~J183039--1002, located within diffuse radio emission and likely associated to a compact radio source.
We use unpublished \emph{Chandra} and \emph{XMM-Newton} observations to localize and identify the X-ray counterpart of this high energy emitter and to characterize its broad-band spectrum. We provide arguments for the identification of this X/gamma-ray source with either a new composite Supernova/PWN system or a background AGN; this second option is however the more convincing on the basis of the source spectral characteristics.

\section{The  \emph{ASCA/INTEGRAL} source AX~J183039--1002}

AX~J183039--1002 was first reported by Sugizaki et al. (2001) as an unidentified and faint X-ray source 
detected by \emph{ASCA} during the Galactic Plane Survey. It was located at RA(J2000) = 18$^{\rm 
h}$30$^{\rm m}$39$^{\rm s}$ and Dec(J2000) = --10\degr02\arcmin42\arcsec\ with a positional uncertainty of 
$\sim$3\arcmin. Its soft X-ray spectrum is well described ($\chi^{2}/\nu=4.9/10$) by an absorbed flat 
power law with $\Gamma =0.04^{+0.77}_{-0.69}$, $N_{\rm H}=\left(3.1^{+3.4}_{-2.6}\right) \times 10^{22}$ cm$^{-2}$ and an unabsorbed
0.7--10 keV flux of $\sim$$2.3\times10^{-12}$ erg cm$^{-2}$ s$^{-1}$.

AX~J183039--1002 was then reported as an IBIS source both by Bird et al. (2007) and Krivonos et al. (2007); 
here we use data collected in the 3rd IBIS survey (Bird et al. 2007), which consists of all exposures from 
the beginning of the mission (November 2002) up to April 2006. The total exposure on this region is $\sim 
$1.78 Ms. ISGRI images for each available pointing were generated in various energy bands using the ISDC 
offline scientific analysis software version 5.1 (OSA~5.1; Goldwurm et al. 2003). The individual images 
were then combined to produce mosaics of the sky region of interest here to enhance the detection 
significance using the system described in detail by Bird et al. (2004, 2007). 
A clear excess is observed in the IBIS map with a significance of 
$\sim$8$\sigma$ at a position corresponding to RA(J2000)=18$^{\rm h}$30$^{\rm m}$39\fs8 and 
Dec(J2000)=--10\degr00\arcmin25\farcs2 and with an associated uncertainty of 4.8\arcmin\ (90\% confidence 
level). 
Our position is compatible with that reported by Krivonos et al. (2007) (note that their 2.1\arcmin\ error radius corresponds to 68\% confidence level). 
Figure 1 shows the 20--40 keV image of the region surrounding AX~J183039--1002 with superimposed both the \emph{INTEGRAL} and \emph{ASCA} error circles. 
The IBIS/ISGRI spectrum was obtained following usual procedures: fluxes were extracted from the location of the source in 10 narrow energy bands over the 17--100 keV range for all available pointings; a spectral \texttt{pha} file was then made by taking the weighted mean of the light curve obtained in each band. 
An appropriately re-binned \texttt{rmf} file was also produced from the standard IBIS spectral response file to match the \texttt{pha} file energy bins. 
Here and in the following, spectral analysis was performed with XSPEC v.11.3.2 package and errors are 
quoted at 90\% confidence level for one interesting parameter ($\Delta\chi^{2}=2.71$). 
A simple power law provides a good fit to the IBIS data ($\chi^2/\nu = 7.6/8$) and a photon index $\Gamma=3.1^{+0.9}_{-0.7}$ (much steeper than the \emph{ASCA} one) combined to an observed 20--100 keV flux of 8.6 $\times$ 10$^{-11}$ erg cm$^{-2}$ s$^{-1}$.

\begin{figure} 
\centering
\resizebox{\hsize}{!}{ \includegraphics{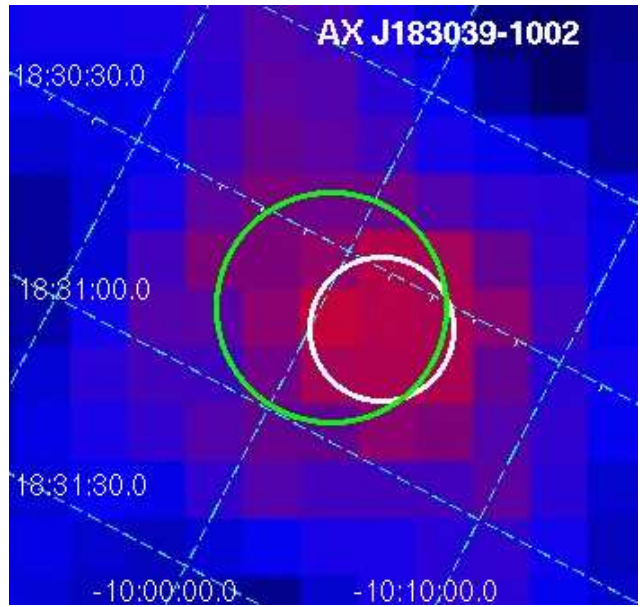} }
\caption {IBIS/ISGRI 20--40 keV image of AX J183039--1002.
The green and white circles correspond to the \emph{INTEGRAL}/IBIS and \emph{ASCA}/GIS error boxes.} 
\label{fig1}
\end{figure}

\section{Radio counterpart}

The Multi-Array Galactic Plane Imaging Survey (MAGPIS, Helfand et al., 2006) has mapped the region 
of interest here with a 20 cm sensitivity of 1--2 mJy and detected a compact source surrounded by diffuse 
extended emission (see Figure 2 which is a MAGPIS extract). 
The compact source (G21.63201--0.00682) which is located at RA(J2000)=18$^{\rm h}$30$^{\rm m}$38\fs29 and Dec(J2000)=--10\degr02\arcmin46\farcs4 (5\arcsec\ uncertainty) has an integrated 20 cm flux density of $5.13\pm0.26$ mJy, and is also detected at 6 cm with an integrated flux density of $1.89\pm0.18$ mJy (White, Becker \& Helfand, 2005). 
The diffuse emission (G21.6417+0.0000) around this discrete source has an integrated flux of 1.54 Jy at 20 cm and a diameter of 2.8\arcmin.

Comparison with mid-infrared images obtained with MSX (Price et al., 2001) allows us to discriminate between 
thermal and non-thermal sources: indeed high values of the radio-to-IR flux ratio (20 cm versus 20 $\mu$m, 
where the source is generally not detected in the MSX) generally indicate non thermal radio emission such as is 
produced in a supernova remnant (SNR, see, e.g., F\"urst et al. 1987, Green 1989). 
This evidence together with the morphology of the diffuse radio emission (in this case almost a complete shell structure) and the detection of the source in the 90 cm images, led Helfand and co-workers to propose G21.6417+0.0000 as a high probability SNR candidate.

The association of a hard X-ray source coincident with a radio point source and surrounded by a very low 
surface brightness radio shell is suggestive of a situation where a SNR harbors a classic young radio pulsar possibly together with its wind nebula (PWN). 
A displacement of the compact source with respect to the centre of the shell structure is not unusual, since many pulsars are born with high kick velocities which allow them to propagate through the shocked ejecta in the SNR interior (Gaensler and Slane 2006).
Although the distance and age of G21.6417+0.000 are not well known,
 it is clear from its small angular size that it's relatively small and
 young. 
 If the compact source is a PWN, therefore, its offset from the centre of 
 the shell would require an unusually large kick velocity.
A way to know if our discrete radio source is a PWN is by means of its radio spectrum, which is expected to have a flat energy index $0.3 \le \alpha \le 0.1$ (defined by $S_{\nu}\propto\nu^{-\alpha}$, Chevalier 1998): 
comparison between 20 and 6 cm data provides an $\alpha$ value of 0.9, considerably steeper than what generally found in PWN. 
Also, the ratio between radio and X-ray core luminosities is higher ($\sim$2.3$\times$10$^{4}$) than usually found in PWNe (Becker \& Helfand, 1987), 
indicating either a highly unusual PWN or an alternative nature for AX~J183039--1002, for example that of a 
background AGN. Indeed the radio spectrum of 0.9 is typical of an active galaxy, although this scenario 
is disregarded at first glance for the low probability of a chance alignment between an AGN and the SNR. 
However a more in depth analysis indicates that the probability of such an association is small but not insignificant. The unabsorbed 2--10 
keV X-ray flux of the likely counterpart of AX~J183039--1002 is about 2$\times$10$^{-12}$ erg cm$^{-2}$ 
s$^{-1}$, and from the Log\,$N$--Log\,$S$ relation (La Franca et al., 2005) we estimate a density for such 
bright AGNs of about 0.1 sources per square degree. The SNR has a radius of about 2.8\arcmin, i.e. an area 
of 6.15 arcmin$^2$. We have then a mean value of about 2$\times$10$^{-4}$ AGNs inside the SNR. However, we have observed in all the sky about 200--300 SNRs (for 
example Green's catalog reports 265 sources, see \texttt{http://www.mrao.cam.ac.uk/surveys/snrs/}). 
Assuming that all of these have the same area of G21.6517+0.000, we can next estimate the probability of finding at least one chance association with an AGN to be about 5\%. 
This probability is small but not insignificant. 
Moreover, many SNRs have a radius much larger than 2.8\arcmin, therefore the ``true'' chance association probability should be considerably greater than $\sim$5\%.

\begin{figure}
\centering
\resizebox{\hsize}{!}{ \includegraphics{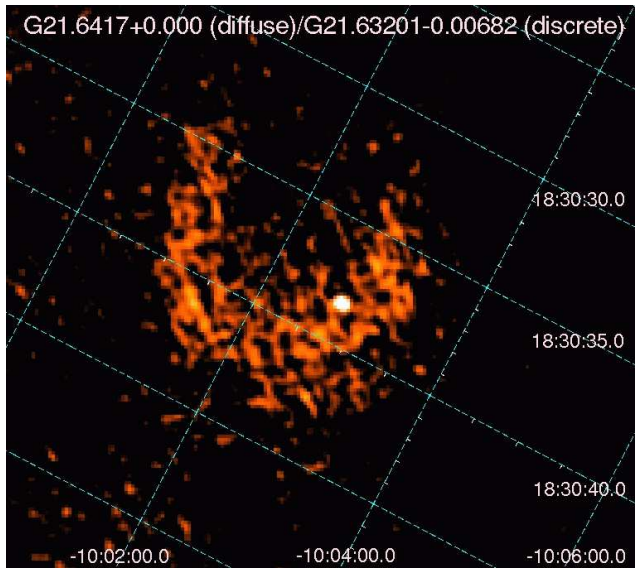} }
\caption{MAGPIS (20 cm) image of AX J183039--1002, showing both the diffuse emission and the compact radio source.
}  
\label{fig2}
\end{figure}

\section{\emph{Chandra} observation}

On February 19, 2006 \emph{Chandra} carried out an observation of the sky region containing AX J183039--1002. 
Data reduction was performed with CIAO v.3.4 and CALDB v.3.2.4 to apply the latest gain corrections. 
Subsequent filtering on event grade and exclusion of high background times resulted in a total exposure of 19.2 ks. 
In order to locate the source of hard X/gamma-ray emission associated to the \emph{ASCA}/IBIS object, we concentrate  the \emph{Chandra} image analysis to the 3--10 keV band: within the smaller \emph{ASCA} error box we detect 2 objects above a 3$\sigma$ detection threshold; their positions, detection significance and count rates are reported in Table~1.

\begin{table}
\centering
\caption{\emph{Chandra} X-ray detections.}\label{tab1}
\begin{tabular}{ccccc}
\hline
 Source & RA (J2000)$^{\dagger}$ & Dec (J2000)$^{\dagger}$ & S/N & Count Rate$^{\ddagger}$ \\
\hline
N1 & 18 30 38.3 & -10 02 47.1  & 27  &  $38.90\pm1.50$     \\
N2 & 18 30 37.1 & -10 02 45.4  & 4.7 &   $1.25\pm0.28$    \\
\hline
\end{tabular}
\begin{flushleft}
$^{\dagger}$  {\footnotesize 1$\sigma$ uncertainty on \emph{Chandra} positions is 0.6\arcsec.} \\
$^{\ddagger}$ {\footnotesize In units of 10$^{-3}$ counts s$^{-1}$ in the 3--10 keV energy band.}
\end{flushleft}
\end{table}

 Figure 3 is a cut-out of the 3--10 keV \emph{Chandra} image, showing the 2 objects detected within 
the \emph{ASCA} error box, together with the location of the compact MAGPIS radio source G21.63201--0.00682. 
It is evident from the figure that the \emph{Chandra} observation further confirms the association of source N1 
with the compact MAGPIS radio object and, purely for its brightness, also with the \emph{ASCA}/IBIS 
source.
The reduced positional uncertainty does not contain any optical/infrared counterpart for this 
\emph{Chandra} source in the HEASARC archive and even in the DSS infrared image (see Sect. 6 for a discussion on the expected optical extinction).

To look for the possible presence of a PWN around source N1, we have also compared the \emph{Chandra} 
PSF against that of the source but have found no evidence for it: object N1 is seen as point-like by 
\emph{Chandra} implying that a PWN is not present or, if present, is extremely compact.

The CIAO script \texttt{psextract} was then used to generate the spectrum, with appropriate background and 
response files, of both source N1 and N2; spectra were extracted using a radius of 
5\farcs6, while background files were generated using a circular region of radius 100\arcsec. 
With a count rate of 0.119 counts/frame, the pile-up fraction is insignificant at less than 2\%.
In all our fitting procedures we used a Galactic column density, which in the direction of AX~J183039--1002 is 1.52 $\times$ 
10$^{22}$ cm$^{-2}$ (Dickey \& Lockman, 1990).

The spectrum of source N2 is very soft with very few counts above 3 keV; when compared to the IBIS 
spectrum we have to assume a cross-calibration constant $\ge 4\times10^{4}$, implying 
that the extrapolation of the Chandra data predicts an IBIS flux far below that observed,
a further indication that this source is not the counterpart of the 
\emph{ASCA}/IBIS object. 

The spectrum of source N1 is instead very hard as a simple power law model 
provides a flat spectrum ($\Gamma=-0.86$) and an unacceptable fit ($\chi^{2}/\nu = 87.8/45$). Residuals to 
this model show some curvature possibly due to intrinsic absorption. Addition of this extra component 
provides a fit improvement  ($\chi^{2}/\nu = 40/45$), which is significant at the 99.99\% confidence level, a steeper spectrum 
($\Gamma = 1.01^{+0.57}_{-0.38}$), a column density of $N_{\rm H}=\left(11.4^{+3.8}_{-2.9}\right) \times 10^{22}$ 
cm$^{-2}$, and an unabsorbed 0.7--10 keV flux of $\left(3.2^{+0.4}_{-0.9}\right)\times 10^{-12}$ erg cm$^{-2}$ 
s$^{-1}$. 
The \emph{Chandra} spectrum is compatible within errors with the \emph{ASCA} one.
The spectrum is quite flat (as found previously) and the column density exceeds largely the Galactic $N_{\rm H}$ value: 
it points to a source able to emit from a few keV up to $\sim$100 keV, constant in time 
and, as underlined before, able to produce emission also at radio frequencies.

\begin{figure}
\resizebox{\hsize}{!}{ \includegraphics{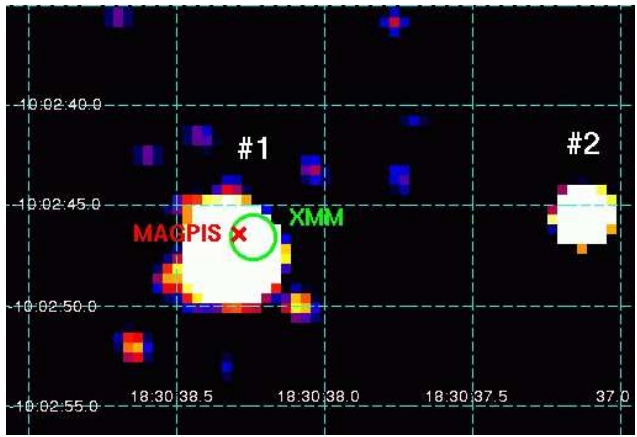} }
\caption {Cut-out of the \emph{Chandra} 3--10 keV band image showing the 2 objects detected within the 
\emph{ASCA} error 
box, together with the \emph{XMM-Newton} 1.06\arcsec\ 1$\sigma$ uncertainty as well as the location of the compact MAGPIS radio object.}
\label{fig3}
\end{figure}

\section{\emph{XMM-Newton} observation}

The \emph{Chandra} N1 source has also been detected during the
XMM Galactic Plane survey (Hands et al., 2004)  
as XGPS--I~J183038--100248.
It is also reported in the 2nd Serendipitous Source Catalogue as 2XMM~J183038.2--100246 
(Watson et al., 2007, \texttt{http://xmmssc-www.star.le.ac.uk}); 
the source 0.2--12 keV flux is $\sim$2$\times$10$^{-12}$ erg cm$^{-2}$ s$^{-1}$ (i.e. compatible with the \emph{ASCA/Chandra} flux), with almost all counts coming from the hardest X-ray band (4.5--12 keV),
again an indication of a flat 2--10 keV spectrum.

We have then analyzed the spectrum of 2XMM~J183038.2--100246, using its pipelined and reduced 2XMM catalogue data products available at the HEADAS archive\footnote{\texttt{http://heasarc.gsfc.nasa.gov/W3Browse/xmm-newton/ xmmssc.html}}. 
This observation was performed on September 13, 2003, for a total of 4335 seconds of exposure time with the EPIC-pn instrument.

\begin{figure}
\centering
\resizebox{\hsize}{!}{ \includegraphics[angle=-90]{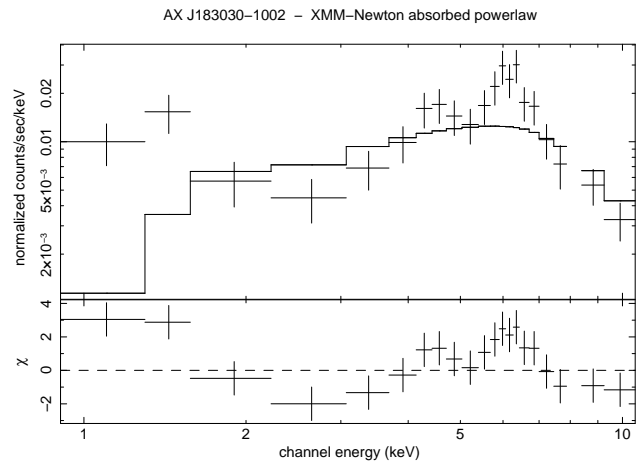} }
\caption{\emph{XMM-Newton} spectrum of AX J183039--1002 fitted with an absorbed power law: 
note the systematic residuals.}
\label{XMMpowerlaw}
\end{figure}

The source was constant during the entire observation and its spectrum is poorly described by a simple absorbed power law ($\chi^{2}/\nu = 62/19$), at variance with that obtained by \emph{ASCA} and \emph{Chandra}.
Excess counts are visible at low energies and at around 6.4 keV (Figure \ref{XMMpowerlaw}).
A good description of the data ($\chi^{2}/\nu = 12/14$) is achieved with a model consisting of a blackbody ($kT = 0.20\pm0.02$ keV), an absorbed power law ($N_{\rm H} = (8.8\pm4.0)\times10^{22}$ cm$^{-2}$; $\Gamma =  0.6\pm0.5$) and an iron $K_\alpha$ line. 
Both these two extra features are required by \emph{XMM-Newton} data, with a significance of 99.99\% for the blackbody component and 99.6\% for the line.
This line is very little redshifted ($z\le0.08$) and has an equivalent width of $\sim$1 keV ($\sigma = 0.4 \pm 0.2$ keV). 
Considering the errors on the line width and normalization given by the low statistics of the data,
the EW could be significantly lower than 1 keV. If present, the broad line may be due to a blend of narrow lines (as often seen in Compton-thick AGNs, e.g. Matt et al. 2004).
Note that \emph{Chandra} does not clearly detect the line. This is not surprising, 
since the effective area of ACIS in the Fe-K band is so small.

\begin{figure}
\centering
\resizebox{\hsize}{!}{ \includegraphics[angle=-90]{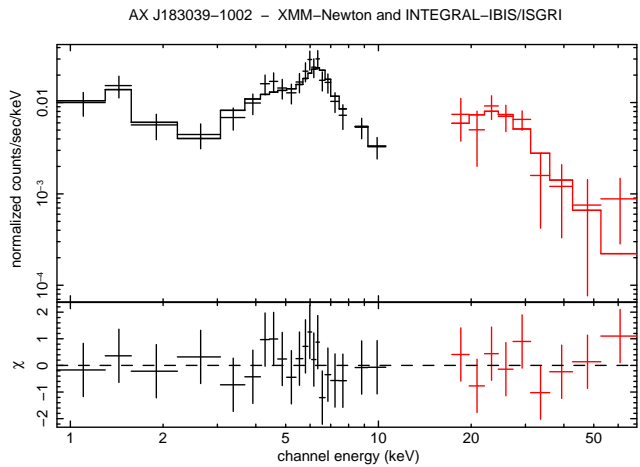} }
\caption{Broad-band spectrum of AX J183039--1002 fitted with a low energy blackbody, a redshifted iron line and an absorbed broken power law: 
\emph{XMM-Newton} data to the left and IBIS data to the right.}
\label{fig4}
\end{figure}

Finally, we combined  the \emph{XMM-Newton} data with the \emph{INTEGRAL} one to obtain a broad-band spectrum over the 1--200 keV energy band. 
The model used to fit the \emph{XMM-Newton} spectrum gives residuals in the IBIS data suggesting the presence of some curvature or a cut-off above 10 keV, which is immediately obvious also from the comparison between the X and $\gamma$-ray photon indices. 
To describe the broad-band source spectrum and in view of the low statistical quality of the data we have 
therefore replaced the absorbed power law component with other simple models, like an absorbed broken power law or an absorbed cut-off power law; 
we have further fixed the cross-calibration constant $C$ between \emph{XMM-Newton} and IBIS to 1 
in view of the non-variability of the source flux in the keV band given by comparing the \emph{ASCA}, \emph{Chandra} and \emph{XMM-Newton} fluxes. 
We emphasize that the fit does not change considerably leaving $C$ free to vary and in any case the large uncertainty on this parameter reaches as a lower boundary $C=1$. 
The first model gives the same parameters found when using \emph{XMM-Newton} data alone, a 
break at $26\pm5$ keV where the hard power law steepens to $\Gamma > 2.8$ ($\chi^{2}/\nu = 12/21$). 
This fit is shown in Figure \ref{fig4}.
The second model, apart from the other parameters that don't change significantly, has $\Gamma=0.0\pm0.5$ with a cut-off energy at $14\pm5$ keV ($\chi^{2}/\nu = 18/22$). 

The two fits are equally acceptable and can be used to describe in a phenomenological way the source spectrum, 
i.e a low-energy blackbody together with an absorbed flat power law steepening at around 10--30 keV and a low-redshift iron line.

\section{What type of source is AX~J183039--1002?}

AX~J183039--1002 is certainly a broad-band emitter over the 1--100 keV band; it is also a weak and compact 
radio source sitting inside or behind a possible supernova remnant. 
No optical source is present within the restricted \emph{Chandra} error box down to a limiting $B$ magnitude of $\sim$20--21; this suggests either that the source is weak or heavily obscured. 
Outside the \emph{Chandra} positional uncertainty we find a DENIS and a 2MASS object at  1.1\arcsec\ and 2.6\arcsec\ angular distance, respectively.
Both have similar near infrared magnitudes ($J = 15.8$ and $K = 13$) and compatible RA and Dec within the respective positional errors so that they can be assumed to be the same source. 
While the DENIS position is compatible within errors with the \emph{Chandra} one, the 2MASS
location is too offset from the X-ray one to claim a match; 
the positional accuracy of the 2MASS catalogue is $\le$1\arcsec, which combined with the Chandra uncertainties provides at most an angular distance of 1.6\arcsec, smaller  than the observed one. 
The mismatch is however not too big and could be explained by a difference in object centroids, due for example to the source shape (as expected in an extended source like a galaxy) or to the source proper motion (as typical of a galactic source).
We therefore assume that the DENIS/2MASS source could be the near-infrared counterpart of
AX~J183039--1002.

The overall morphology of the source suggests two possible scenarios: a PWN associated to the SNR or a 
background AGN. In the first case the source would be a very unusual pulsar/PWN system from the radio and 
X/soft gamma-ray spectral characteristics.
For example, the soft X-ray spectral index of a PWN is usually around 1.5--2 (Kargaltsev \& Pavlov 2008), softer than the \emph{ASCA} and \emph{Chandra} spectra. Furthermore, the photon index $\Gamma > 2.8$ observed in the IBIS band is significantly higher than the usual values for the \emph{INTEGRAL} PWNe, that cluster around $\Gamma \sim 2.2$ (Dean et al. 2008).
The alternative possibility that we are dealing with a magnetar is 
disregarded on the basis of the stable 2--10 keV flux over the years and the shape of the high energy 
emission (typically IBIS sees a hard tail in this type of sources, Kuiper et al. 2006).

In the second, more likely, case we would deal with a heavily absorbed AGN (a type-2 Seyfert?) with an extremely low exponential cut-off energy. 
The radio flux and spectral index would be compatible with the range of values 
found in other \emph{INTEGRAL} selected AGN, which are almost invariably associated with radio sources (Mushotzky 2004). 
In this case, the spatial coincidence between a background AGN and a SNR is fortuitous.
But the more conclusive evidence for the association with a heavily absorbed AGN comes from the \emph{XMM-Newton} spectrum, that shows a strong iron line and is reminescent of a Compton-thick AGN.
We have phenomenologically fitted the source spectrum with simple models, like a blackbody and broken power law.
From a comparison with other similar sources in the literature (e.g. NGC~1068, Matt et al. 1997; see also Maiolino et al. 1998) we can associate the low-energy blackbody-like component with a warm reflector, and the high-energy component as a cold reflector coming from the inner torus. 
The ultimate solution to this puzzle will only come from detailed radio observations: 
hopefully it will not be long before AX~J183039--1002 is definitely properly classified. 
If the heavily absorbed nature of this source is confirmed, this brings to 5 the number of candidate Compton-thick AGN in the third IBIS catalogue, besides the 7 confirmed sources, i.e. about 20\% of the total sample of type-2 active galaxies so far detected (Bird et al. 2007).

\section*{Acknowledgments}
We acknowledge ASI  financial and programmatic support via contracts I/008/07/0 and I/088/06/0.
We also acknowledge the funding by PPARC grant PP/C000714/1.

\label{lastpage}

\begin{thebibliography}{}

\bibitem{Becker87a} Becker, R. H., \& Helfand, D. J. 1987, AJ, 94, 1629

\bibitem{Becker87b} Becker, R. H., \& Helfand, D. J. 1987, ApJ, 316, 660

\bibitem{Bird04} Bird, A. J., Barlow, E. J., Bassani, L., et al. 2004, ApJ, 607, L33

\bibitem{Bird07} Bird, A. J., Malizia, A., Bazzano, A., et al. 2007, ApJS, 170, 175

\bibitem{Chevalier98} Chevalier, R. A., 1998, Mem. S. A. It., 69, 977

\bibitem{Dean08} Dean, A. J., et al., 2008, MNRAS, submitted

\bibitem{Furst87} F\"urst, E., Reich, W., Sofue, Y., 1987, A\&ASS, 71, 63

\bibitem{Gaensler07} Gaensler, B. M. \& Slane, P. O.,  2007, ARA\&A, 44, 17

\bibitem{Green89} Green, D. A., 1989, AJ, 98, 1358

\bibitem{Goldwurm03} Goldwurm, A., David, P., Foschini, L., et al. 2003, A\&A, 411, L223

\bibitem{Hands04} Hands, A. D. P., Warwick, R. S., Watson, M. G., \& Helfand, D. J., 2004, MNRAS, 351, 31

\bibitem{Helfand06} Helfand, D. J., Becker, R. H., White, R. L., Fallon, A., \& Tuttle, S., 2006, AJ, 131, 2525

\bibitem{Kargaltsev08} Kargaltsev, O. \& Pavlov, G. G., 2008, AIP Conference Proceedings, 983, 171

\bibitem{Krivonos07} Krivonos, R., Revnivtsev, M., Lutovinov, A., et al. 2007, A\&A, 475, 775

\bibitem{Kuiper06} Kuiper, L., Hermsen, W., den Hartog, P. R., Collmar, W. 2006, ApJ, 645, 556

\bibitem{LaFranca05} La Franca, F., Fiore, F., Comastri, A., et al., 2005, ApJ, 635, 864

\bibitem{Maiolino98} Maiolino, R., Salvati, M., Bassani, L., et al. 1998, A\&A, 338, 781 

\bibitem{Masetti08} Masetti, N., Mason, E., Morelli, L., et al. 2008, A\&A, 482, 113 

\bibitem{Matt97} Matt, G., Guainazzi, M., Frontera, F., et al. 1997, A\&A, 325, L13 

\bibitem{Matt04} Matt, G., Bianchi, S., Guainazzi, M., Molendi, S., 2004, A\&A,  414, 155

\bibitem{Mushotzky04} Mushotzky, R., ASSL, 308, 53 

\bibitem{Predehl95} Predehl, P. \& Schmitt, J. H. M. M. 1995, A\&A, 293, 889

\bibitem{Price01} Price, S. D., Egan, M. P., Carey, S. J., et al. 2001, AJ, 121, 2819

\bibitem{Stephen06} Stephen, J. B., Bassani, L., Malizia, A., et al. 2006, A\&A, 445, 869

\bibitem{Sugizaki01} Sugizaki, M., Mitsuda, K., Kaneda, H., et al. 2001, ApJS, 134, 77

\bibitem{Ubertini03} Ubertini, P., Lebrun, F., Di Cocco, G., et al. 2003, A\&A, 411, L131

\bibitem{Watson08} Watson, M. G., Schr\"{o}der A. C., Fyfe, D., et al. 2008, astro--ph(0807.1067)

\bibitem{White05} White, R. L., Becker, R. H., \& Helfand, D. J. 2005, AJ, 130, 586

\bibitem{Winkler03} Winkler, C., Gehrels, N., Sch\"{o}nfelder, V., et al. 2003, A\&A, 411, L355


\end{thebibliography}
\end{document}